\documentclass[twocolumn,prl,showpacs,superscriptaddress,floatfix,
nofootinbib]{revtex4-1}
\usepackage{color}
\usepackage{epsfig}
\usepackage{amsmath,bm}
\begin{document}

\title{Neutrino and axion bounds from the globular cluster M5 (NGC 5904)}

\author{N.~Viaux}

\affiliation{Pontificia Universidad Cat{\'o}lica de Chile, Instituto de Astrof{\'\i}sica, Facultad de F{\'\i}sica, Av.\ Vicu{\~n}a
Mackenna 4860, 782-0436 Macul, Santiago, Chile}

\affiliation{Pontificia Universidad Cat\'olica de Chile, Centro de
Astroingenier\'ia, Av.\ Vicu\~na Mackena 4860, 782-0436 Macul,
Santiago, Chile}

\affiliation{The Milky Way Millennium Nucleus, Av.\ Vicu{\~n}a
Mackenna 4860, 782-0436 Macul, Santiago, Chile}

\author{M.~Catelan}

\affiliation{Pontificia Universidad Cat{\'o}lica de Chile, Instituto de Astrof{\'\i}sica, Facultad de F{\'\i}sica, Av.\ Vicu{\~n}a
Mackenna 4860, 782-0436 Macul, Santiago, Chile}

\affiliation{Pontificia Universidad Cat\'olica de Chile, Centro de
Astroingenier\'ia, Av.\ Vicu\~na Mackena 4860, 782-0436 Macul,
Santiago, Chile}

\affiliation{The Milky Way Millennium Nucleus, Av.\ Vicu{\~n}a
Mackenna 4860, 782-0436 Macul, Santiago, Chile}

\author{P.~B.~Stetson}

\affiliation{National Research Council, 5071 West Saanich Road,
Victoria, BC V9E 2E7, Canada}

\author{G.~G.~Raffelt}

\affiliation{Max-Planck-Institut f\"ur Physik
(Werner-Heisenberg-Institut), F\"ohringer Ring 6, 80805 M\"unchen,
Germany}

\author{J.~Redondo}

\affiliation{Arnold Sommerfeld Center,
Ludwig-Maximilians-University, Theresienstr.~37, 80333 M\"unchen,
Germany}

\affiliation{Max-Planck-Institut f\"ur Physik
(Werner-Heisenberg-Institut), F\"ohringer Ring 6, 80805 M\"unchen,
Germany}

\author{A.~A.~R.~Valcarce}
\affiliation{Universidade Federal do Rio Grande do Norte, Depto.\ de
F\'{i}sica, 59072-970 Natal, RN, Brazil}

\author{A.~Weiss}
\affiliation{Max-Planck-Institut f\"ur Astrophysik,
Karl-Schwarzschild-Str.~1, 85748 Garching, Germany}

\date{5 November 2013}

%%%%%%%%%%%%%%%%%%%%%%%%%%%%%%%%%%%%%%%%%%%%%%%%%%%%%%%%%%%%%%%%%%%

\begin{abstract}
The red-giant branch (RGB) in globular clusters is extended to larger
brightness if the degenerate helium core loses too much energy in
``dark channels.'' Based on a large set of archival observations, we
provide high-precision photometry for the Galactic globular cluster
M5 (NGC 5904), allowing for a detailed comparison between the
observed tip of the RGB with predictions based on contemporary
stellar evolution theory. In particular, we derive 95\% confidence
limits of $g_{ae}<4.3\times10^{-13}$ on the axion-electron coupling
and $\mu_\nu<4.5\times10^{-12}\,\mu_{\rm B}$ (Bohr magneton $\mu_{\rm
B}=e/2m_e$) on a neutrino dipole moment, based on a detailed analysis
of statistical and systematic uncertainties. The cluster distance is
the single largest source of uncertainty and can be improved in the
future.
\end{abstract}


\maketitle

%%%%%%%%%%%%%%%%%%%%%%%%%%%%%%%%%%%%%%%%%%%%%%%%%%%%%%%%%%%%%%%%%%%%%%
% Introduction
%%%%%%%%%%%%%%%%%%%%%%%%%%%%%%%%%%%%%%%%%%%%%%%%%%%%%%%%%%%%%%%%%%%%%%

{\em Introduction.}---Astrophysics and cosmology provide us with
powerful arguments to constrain the properties of elementary
particles. The ``heavenly laboratories'' are complementary to
terrestrial experiments, notably at the low-energy frontier of
particle physics, which includes the physics of neutrinos and other
weakly interacting low-mass particles such as the hypothetical
axion. In particular, stars would lose energy by emitting such
particles in addition to standard neutrinos, leading to potentially
observable modifications of the properties of individual stars or of
entire stellar populations~\cite{Bernstein:1963qh, Stothers:1970ap,
Dicus:1978fp, Raffelt:1990yz}.

Different types of stars provide information on different particles
or interaction channels because the energy-loss rate of the hot
stellar medium depends on temperature and density in ways determined
by the emission process. For example, low-mass hidden photons are
most significantly constrained by properties of the Sun
\cite{An:2013yfc}. On the other extreme, the neutrino burst duration
of supernova 1987A provides the most restrictive limit on the
axion-nucleon interaction \cite{Raffelt:1987yt}. In many other cases,
evolved low-mass stars---red giants and horizontal branch (HB) stars
in globular clusters (GCs) or white dwarfs (WDs)---supply the most
interesting information \cite{Raffelt:1990pj, Raffelt:1992pi,
Raffelt:1994ry, Raffelt:1989xu, Catelan:1995ba}.

One particularly sensitive observable is the brightness of the tip of
the RGB (TRGB) in GCs \cite{Raffelt:1990pj, Catelan:1995ba,
Raffelt:1994ry, Catelan:1995ba}. Together with other observables such
as the HB brightness, it was found that the core mass at helium
ignition should not exceed its standard value by about 5\%
\cite{Raffelt:1989xu, Catelan:1995ba}. This constraint means that the
energy loss rate should not exceed standard neutrino emission by more
than about a factor of three.

The helium core before ignition is highly degenerate
\cite{Catelan:2009} and neutrinos are primarily emitted by plasmon
decay $\gamma\to\bar\nu\nu$. A sizeable magnetic dipole moment
$\mu_\nu$ would enhance this process \cite{Bernstein:1963qh} and the
TRGB brightness provides the most restrictive $\mu_\nu$ limit to date
\cite{Catelan:1995ba}. Another important constraint is on the
axion-electron coupling $g_{ae}$, where the  most relevant emission
reaction is axio-bremsstrahlung $e+Ze\to Ze+e+a$. It is these cases
that we will re-examine here.

The main motivation for returning to this subject is the enormous
observational progress and especially the newly available, exquisite
GC color-magnitude diagrams (CMDs) that have become available only
recently, both based on ground- and space-based observations
\cite[e.g.,][]{Bergbusch:2009, King:2012}. Likewise, stellar
evolution theory has seen revolutionary progress, for example, by new
opacity and equation-of-state tables. Moreover, in previous studies,
systematic and statistical errors were not analyzed in sufficient
detail to assign clear quantitative confidence levels, preventing a
simple comparison with laboratory results. Our new constraints are
similar to previous astrophysical limits
\cite{Raffelt:1989xu,Raffelt:1992pi,Raffelt:1994ry,Catelan:1995ba} if
the latter are interpreted as $1\sigma$ results. However, we have
used homogeneous observations of a single GC and provide a detailed
error budget.

Technical details, with a focus on the $\mu_\nu$ case, are reported
in a long companion paper \cite{Viaux:2013} (Paper~I). We here
communicate the main points and extend the analysis to the
axion-electron interaction which is of topical interest in view of
some indications for enhanced WD cooling, which we comment on in
more detail below.

%%%%%%%%%%%%%%%%%%%%%%%%%%%%%%%%%%%%%%%%%%%%%%%%%%%%%%%%%%%%%%%%%%%%%%
% Cluster Selection
%%%%%%%%%%%%%%%%%%%%%%%%%%%%%%%%%%%%%%%%%%%%%%%%%%%%%%%%%%%%%%%%%%%%%%

{\em Cluster selection and photometry}.---Among the fully resolved
Milky Way GCs we consider those with an integrated absolute magnitude
$M_V < -8.0$~mag to ensure a well-populated CMD. We restrict
foreground reddening to $E(B-V)\leq 0.1$~mag, also reducing the
possibility of differential reddening. We ensure that the metallicity
is neither too high nor too low, leading to a fairly uniformly
populated HB. Candidates must be sufficiently close that deep,
high-quality photometric data exist. We avoid GCs which seem to have
multiple CMD sequences. These criteria leave us with a short list of
candidates with M5 (NGC 5904)~at the top. It is a well-studied,
fairly massive GC, with $M_V = -8.81$~mag, a moderate metallicity of
${\rm [Fe/H]} = -1.29$, and a foreground reddening of only $E(B-V) =
0.03$~mag. The distance is only a modest 7.5~kpc from the Sun.

We have carried out crowded-field, point-spread
function (PSF) photometry of M5 using the DAOPHOT II/ALLFRAME suite of
programs \cite{Stetson:1987iu}.
Our database was compiled from many sources, including public
archives, following previous works on different GCs \cite{Stetson:2009}.
Current observations consist of 2840 CCD images obtained
during 40 observing runs on 12 telescopes over a
span of 27 years (see Paper~I for details). The resulting CMD
is decontaminated from field stars with a
statistical procedure~\cite{Gallart:2002gf}.

\begin{figure}
\includegraphics[width=0.90\columnwidth]{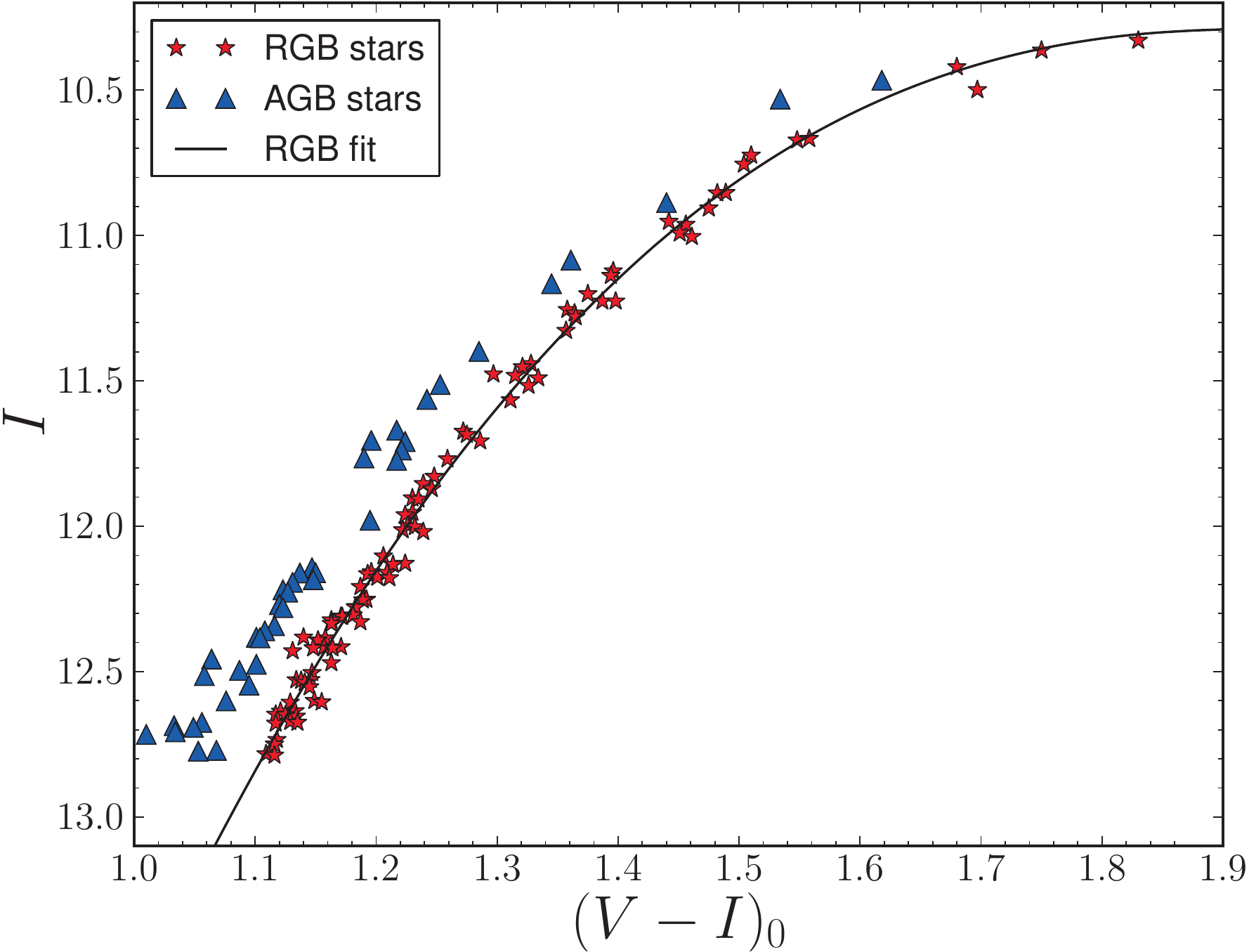}
\caption{Upper CMD for the GC M5, with our RGB and AGB identification
according to color. We also show our empirical
RGB fit function.\label{fig:cmd}}
\end{figure}

The cleaned upper CMD is shown in Fig.~\ref{fig:cmd} together with an
empirical RGB fit function $I=10.289+3.83\,[1.95 - (V-I)_0]^{2.5}$.
We identify stars as belonging to the RGB if their distance from this
line is less than 0.03~mag. Most of the brightest stars are found to
be on the RGB, in agreement with purely statistical expectations.
Another way of discrimination is based on chemical abundance
variation. Other authors also assign the three brightest stars to the
RGB \cite{Sandquist:2004wf, Ivans:2001qp}, except for the second
brightest that could be on the AGB \cite{Sandquist:2004wf}.

The $I$-band magnitudes of the brightest stars are 10.329, 10.363 and
10.420~mag, respectively. These stars are located near the cluster
center, yet the combined error from crowding, completeness and
saturation is probably less than $\pm0.01$~mag. The photometric error
for the brightest star is $\pm0.0057$~mag,~whereas the calibration
error of the $I$-band photometry is not larger than $\pm0.02$~mag
(see Fig.~1 of Ref.~\cite{Stetson:2005fg}). Combining these errors in
quadrature provides a photometric uncertainty of $\sigma_I=0.023$~mag
for the brightest star.

%%%%%%%%%%%%%%%%%%%%%%%%%%%%%%%%%%%%%%%%%%%%%%%%%%%%%%%%%%%%%%%%%%%%%%
% Brightness of TRGB
%%%%%%%%%%%%%%%%%%%%%%%%%%%%%%%%%%%%%%%%%%%%%%%%%%%%%%%%%%%%%%%%%%%%%%

{\em Observed TRGB brightness}.--- The brightest star, whose $I$-band
magnitude we shall denote $I_{1}$, has not yet ignited helium; it
will brighten further, and therefore its current CMD position
provides a lower limit to the TRGB. Using Monte Carlo realizations of
the upper CMD of M5, based on the population shown in
Fig.~\ref{fig:cmd}, we find a statistical TRGB distribution relative
to $I_1$ which has nearly exponential form. On average the TRGB is
$\langle\Delta_{\rm tip}\rangle=0.048$~mag brighter than $I_1$ with
an rms deviation of $\sigma_{\rm tip}=0.058$~mag.

One key ingredient to compare the TRGB brightness with theoretical
predictions~is the distance modulus to M5. Different methods lead to
estimations falling in the range $14.32\leq (m-M)_0\leq14.67$ (see
Table~4 of Ref.~\cite{Coppola:2011yg}) but several of them depend on
HB and RR Lyrae stars whose properties depend on additional cooling
in their RGB progenitors. Therefore, to avoid circular reasoning, we
only use $(m-M)_0=14.45 \pm 0.11$ derived via main-sequence fitting
\cite{Layden:2005fh}, which is unaffected by the exotic energy loss
channels  we discuss here. It is also in excellent agreement with
other distance indicators and already takes into account interstellar
extinction.

We estimate the absolute $I$-band TRGB brightness as $M_{I,\rm TRGB}^{\rm
obs}=I_1-\langle \Delta_{\rm tip}\rangle -(m-M)_0$, i.e.,
\begin{equation} \label{eq:TRGBobs}
M_{I,\rm TRGB}^{\rm obs}=-4.17\pm 0.13~{\rm mag}\,,
\end{equation}
where we have added the errors in quadrature. The uncertainty
derives almost entirely from the distance.

%%%%%%%%%%%%%%%%%%%%%%%%%%%%%%%%%%%%%%%%%%%%%%%%%%%%%%%%%%%%%%%%%%%%%%
% Prediction
%%%%%%%%%%%%%%%%%%%%%%%%%%%%%%%%%%%%%%%%%%%%%%%%%%%%%%%%%%%%%%%%%%%%%%

{\em Predicted TRGB brightness}.---To predict $M_{I,\rm TRGB}$ we
use the Princeton-Goddard-PUC (PGPUC) code \cite{Valcarce:2012vp} to
calculate evolutionary sequences up to the point of He ignition,
implementing varying amounts of $\mu_\nu$ or axion energy losses.
Our benchmark tracks use $M=0.82\,M_{\odot}$ without mass loss on
the RGB, $Y=0.245$, $Z=0.00136$, and $[\alpha/{\rm Fe}]=+0.30$ to
capture the best estimates for the stellar properties in M5. To
compare with observational data, we transform the luminosity into
$I$-band absolute brightness using the bolometric correction (BC) of
Worthey and Lee \cite{Worthey:2006gj}.

The dominant neutrino emission process on the RGB evolution is
plasmon decay for which PGPUC uses the analytic approximation
formulas of Haft et al.\ \cite{Haft:1993jt}. To incorporate $\mu_\nu$
effects, we scale this rate by the prescription given in Eqs.~(9) and
(10) of Ref.~\cite{Raffelt:1992pi}. The axion-electron interaction,
to be discussed in more detail below, allows for photo-production
(Compton scattering) $\gamma+e\to e+a$ and bremsstrahlung $e+(Z,A)\to
(Z,A)+e+a$ and $e+e\to e+e+a$. We take the energy-loss rate from
Ref.~\cite{Raffelt:1994ry}, but extend their calculation to include
all chemical elements as scattering targets, not helium alone.

The simulated TRGB brightness for the neutrino case can be expressed
in terms of simple analytic fit formulas as $M^0_{I,{\rm TRGB}}=
-4.03-0.23\,[(\mu_{12}^2+0.64)^{0.5}-0.80-0.18\,\mu_{12}^{1.5}]$,
where $\mu_{12}=\mu {_\nu}/10^{-12}\mu_{\rm B}$. For axions the
corresponding result is
$-4.03-0.25\,[(g_{13}^2+0.93)^{0.5}-0.96-0.17\,g_{13}^{1.5}]$, where
$g_{13}=g_{ae}/10^{-13}$ and $g_{ae}$ is the dimensionless
axion-electron Yukawa coupling constant.

These predictions are affected by a number of systematic
uncertainties of stellar evolution theory detailed in Paper~I. Many
of them influence $M_{I,{\rm TRGB}}$ by less than $\pm0.01$~mag and
deserve no further mention here. Larger uncertainties derive from
the helium abundance ($\pm0.010$~mag), conductive opacity
($\pm0.016$~mag), nuclear reaction rates ($\pm0.019$~mag), screening
effects ($\pm0.011$~mag), and standard neutrino emission rates
($\pm0.013$~mag).

More important is the question of mass loss on the RGB. Based on the
HB properties in M5, we argue in Paper~I that stars lose between
0.12 and $0.28\,M_\odot$. The impact on $M_{I,{\rm TRGB}}$ is not
monotonic in this interval, leading to a shift relative to the
no-mass-loss baseline case of between $+0.022$ and $+0.035$~mag.

The uncertainty of the equation of state (EoS) has a similar impact.
PGPUC uses FreeEOS (see Table~2 of Ref.~\cite{Valcarce:2012vp}),
while other codes use other prescriptions. To study the impact of EoS
variations we use the GARSTEC stellar evolution code
\cite{Weiss:2008} with 8 different EoS prescriptions. For FreeEOS and
all other parameters identical to our PGPUC baseline case, the TRGB
is found 0.05~mag brighter. The internal GARSTEC spread of EoS cases
is $-0.0045$ to $+0.0242$~mag in $M_{I,\rm TRGB}$.

The largest theoretical uncertainty derives from the treatment of
convection. PGPUC uses the mixing-length theory (MLT) where the
mixing-length parameter $\alpha_{\rm MLT}$ of convection theory is
chosen to reproduce the Sun. In this way one achieves a quite
satisfactory match of the CMDs of GCs over a wide range of
metallicities, suggesting an uncertainty of $\pm 0.1$ in $\alpha_{\rm
MLT}$. In addition, an uncertainty due to the calibration of
$\alpha_{\rm MLT}$ arises. Depending on the inclusion of atomic
diffusion, another shift of $\pm 0.1$ in $\alpha_{\rm MLT}$~is
conceivable. Overall we adopt an uncertainty of $\pm 0.2$ in
$\alpha_{\rm MLT}$, corresponding to a brightness uncertainty of
$\mp0.056$~mag.

The largest uncertainty in the comparison between theory and
observations comes from the color transformations and BC. Worthey and
Lee \cite{Worthey:2006gj} provide explicit error estimates for their
BC which depends on the luminosity and temperature of the star and
hence on the TRGB locus. For the neutrino case, these results suggest
an error of $\sigma_{\rm BC}=(0.08+0.013\,\mu_{12})~{\rm mag}$. This
uncertainty is considerably larger than the spread of BC values
derived from the prescriptions of other authors. The corresponding
axion result is $\sigma_{\rm BC}=(0.08+0.02\, g_{13})~{\rm mag}$.

\begin{figure}
\includegraphics[width=0.90\columnwidth]{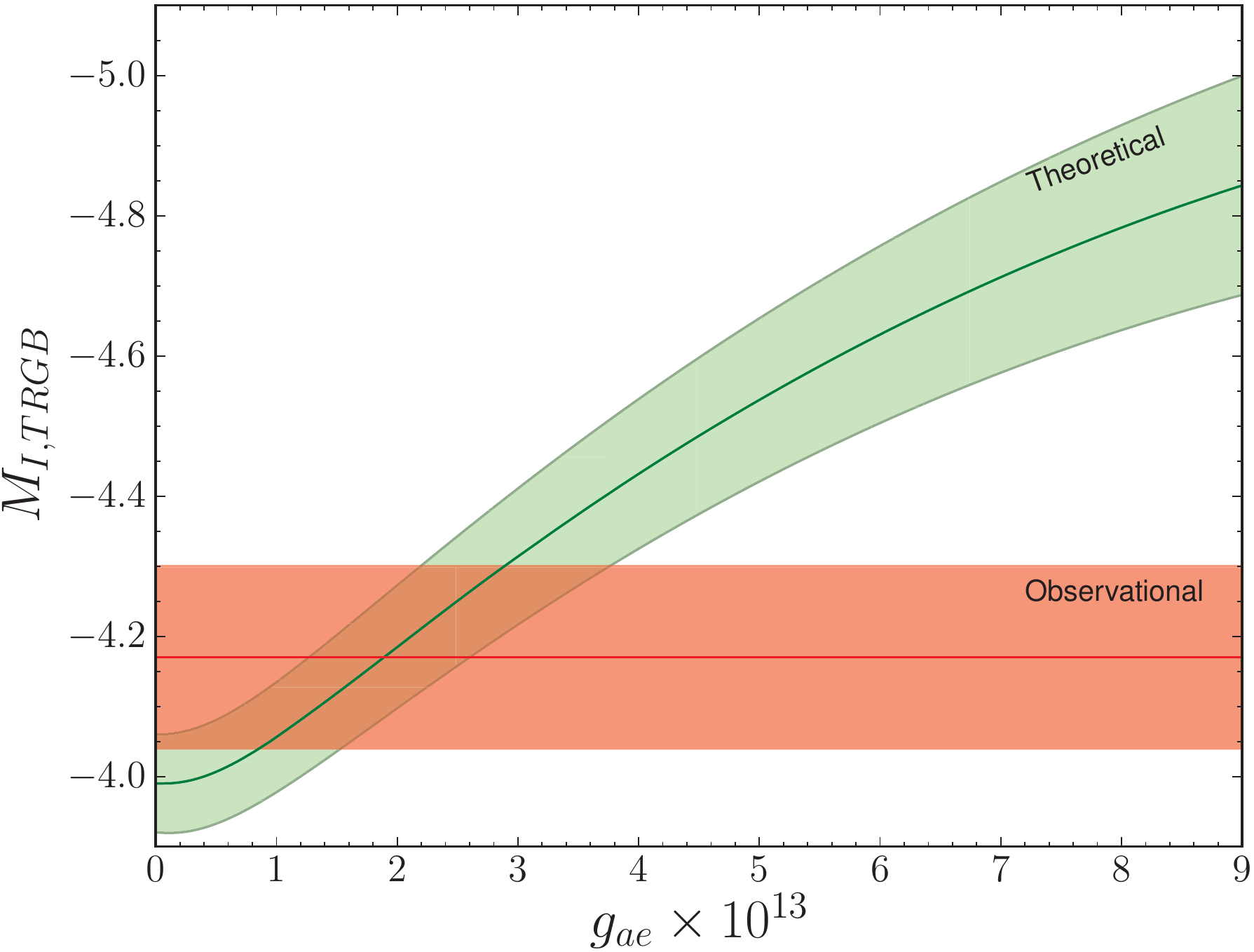}
\caption{Absolute $I$-band brightness of TRGB in cluster M5. {\em Red band:}
Observations with $1\sigma$ error, dominated by distance.
{\em Green band:} Theoretical prediction, depending on the axion-electron
coupling, with $1\sigma$ systematic error, dominated by the bolometric
correction.\label{fig:brightness}}
\end{figure}

All of these uncertainties are systematic (not statistical) and are
our best estimates of the maximum error. The associated probability
distributions are in most cases completely unknown, so we make the
simplest possible choice and use top-hat, flat probability
distributions in the given ranges of $M_{I,{\rm TRGB}}$
modifications. Convolving all of these distributions leads to a
Gaussian distribution with mean 0.039~mag, i.e., $M^{\rm
theory}_{I,{\rm TRGB}}=M^{0}_{I,{\rm TRGB}}+0.039$, and standard
deviation $\sigma_{\rm
theory}=[0.039^2+(0.046+0.0075\mu_{12})^2]^{0.5}$~mag. For axions,
this result is $\sigma_{\rm theory}=[0.039^2+(0.046+0.012
g_{13})^2]^{0.5}$~mag shown as a green band in
Fig.~\ref{fig:brightness}. (A similar figure for the $\mu_\nu$ case
is shown in Paper~I.)

Within the uncertainties, the observed and predicted TRGB brightness
agrees without novel cooling effects. To derive bounds on $\mu_\nu$
and $g_{ae}$ we combine the observational and theoretical errors in
quadrature. Integrating the combined probability distribution from
$\mu_{12}=0$ or $g_{ae}=0$ to the limiting value, we find
\begin{eqnarray}\label{eq:constraints}
\mu_{\nu} &<&2.6~(4.5) \times 10^{-12}\mu_{\rm B}
\nonumber\\
g_{ae} &<&2.6~(4.3)\times 10^{-13}
\end{eqnarray}
at the 68\%~(95\%) CL, respectively.

%%%%%%%%%%%%%%%%%%%%%%%%%%%%%%%%%%%%%%%%%%%%%%%%%%%%%%%%%%%%%%%%%%%%%%
% Axion bounds
%%%%%%%%%%%%%%%%%%%%%%%%%%%%%%%%%%%%%%%%%%%%%%%%%%%%%%%%%%%%%%%%%%%%%%

{\em The axion-electron coupling}.---Axions are hypothetical
pseudoscalar particles that must exist if the Peccei-Quinn mechanism
is the correct explanation for CP conservation in QCD
\cite{Peccei:2006as, Kim:2008hd, Kawasaki:2013ae}. Their properties
are governed primarily by an energy scale, $f_a$, the Peccei-Quinn
scale or axion decay constant. Their mass arises from mixing with the
$\pi^0$, $\eta$ and $\eta'$ mesons and is found to be $m_a=(m_\pi
f_\pi/f_a)\sqrt{z}/(1+z)\sim 6~{\rm meV}\,(10^9~{\rm GeV}/f_a) $ in
terms of the pion mass $m_\pi=135$~MeV, pion decay constant
$f_\pi=93$~MeV, and up/down quark mass ratio $z=m_u/m_d=0.38$--0.58
\cite{Beringer:1900zz}. One generic axion property is its two-photon
vertex that allows for production in stars by the Primakoff process
\cite{Dicus:1978fp, Raffelt:1985nk} and for solar axion searches by
the reverse process \cite{Sikivie:1983ip, vanBibber:1988ge,
Lazarus:1992ry, Moriyama:1998kd, Zioutas:2004hi}. The helium-burning
lifetime of HB stars in GCs \cite{Raffelt:2006cw} as well as the
existence of the blue-loop phase in massive stars
\cite{Friedland:2012hj} provide a limit for typical axion models,
corresponding to $m_a \lesssim 0.3$ eV, which is more stringent than
constraints based on the helium flash in low-mass stars, as discussed
for instance in Refs.~\cite{Raffelt:1987yu, Catelan:1995ba}.

In addition, axions can interact with electrons with a vertex of the
form $C_e\overline\psi_e\gamma^\mu\gamma_5\psi_e\partial_\mu a/2f_a$,
where $C_e$ is a model-dependent coefficient and one usually defines
the dimensionless Yukawa coupling $g_{ae}=C_e m_e/f_a$. A benchmark
case is the DFSZ model \cite{Dine:1981rt} where explicitly
$g_{ae}=\frac{1}{3}\,\cos^2(\beta)\,m_e/f_a$ and $\tan\beta$ is the
ratio between two Higgs-field expectation values. Conversely, this
implies $\widetilde m_a/{\rm meV}=g_{ae}/2.8\times10^{-14}$ where we
have defined $\widetilde m_a=m_a \cos^2\beta$. Our limit on $g_{ae}$
from the TRGB in M5 then implies {$\widetilde m_a<9.3~(15.4)~{\rm
meV}$ at the 68\% (95\%) CL, respectively.

WDs also emit axions efficiently by bremsstrahlung, and the WD
luminosity function allows one to set restrictive limits
\cite{Raffelt:1985nj, Isern:1992, Isern:2008nt, Miller:2013}. In
particular, Isern et al.\ \cite{Isern:2008nt} find that a small
amount of axion cooling, corresponding to $\widetilde m_a\sim 5~{\rm
meV}$, slightly improves the overall fit. On the other hand, a more
consistent implementation of axion cooling reveals $\widetilde
m_a<8~{\rm meV}$ at 95\% CL \cite{Miller:2013}.

The WD cooling speed can also be tested by the period decrease of
pulsating WDs (ZZ Ceti stars). The well-studied case of G117--B15A
shows a decrease of its 215~s period {at a rate of $(4.19 \pm
0.73)\times 10^{-15}~{\rm s/s}$, and requires additional cooling
corresponding to $\widetilde m_a$ of 15--20~meV \cite{Isern:1992,
Isern:2010wz, Corsico:2012ki}. The star R548 shows a similar effect
where additional cooling is required at about 95\% CL
\cite{Corsico:2012sh}. The axion limits from the WD luminosity
function and TRGB brightness exclude strong axion cooling of
pulsating WDs---the apparent period decrease may be caused by other
effects or systematic uncertainties.

Still, the tantalizing possibility remains that axions with meV-range
masses could exist and then play an important role for the cooling of
WDs and neutron stars. If so, core-collapse SNe would emit a
significant fraction of their energy in axions and produce a cosmic
diffuse supernova axion background (DSAB) \cite{Raffelt:2011ft}.

%%%%%%%%%%%%%%%%%%%%%%%%%%%%%%%%%%%%%%%%%%%%%%%%%%%%%%%%%%%%%%%%%%%%%%
% Conclusions
%%%%%%%%%%%%%%%%%%%%%%%%%%%%%%%%%%%%%%%%%%%%%%%%%%%%%%%%%%%%%%%%%%%%%%

{\em Conclusions}.---The observed and predicted $I$-band brightness
of the TRGB in M5 agree reasonably well within uncertainties,
although the agreement would improve with a small amount of extra
cooling that slightly postpones helium ignition. We have implemented
additional cooling by plasmon decay which is enhanced by a neutrino
magnetic dipole moment, $\mu_\nu$, and by axion emission in terms of
the axion-electron Yukawa coupling $g_{ae}$. After adding statistical
and systematic uncertainties in quadrature, we find the 95\% CL
constraints $\mu_\nu<4.5\times10^{-12}\,\mu_{\rm B}$ and
$g_{ae}<4.3\times10^{-13}$. These are comparable to similar
astrophysical bounds in the literature, but are now based on a single
GC and a detailed error budget that has allowed for a reasonably
quantified confidence level. Both limits correspond to $\Delta M_{\rm
c}< 0.047\, M_\odot$ for the non-standard core-mass increase at
helium ignition.

Our limits have not improved as much as one might have hoped because
observations and predictions would agree better with a small amount
of extra cooling, although this effect is not significant within the
uncertainties. Still, it is noteworthy that the WD luminosity
function and period decrease of ZZ~Ceti stars also mildly point to
extra cooling. None of these cases have fluctuated in the opposite
direction of suggesting reduced standard cooling. So perhaps there is
an unrecognized common systematic issue with all of these cases.

Our new TRGB comparison between theory and observations can be
improved in the future because our single largest source of
uncertainty is the cluster distance, which should be improved by the
upcoming GAIA mission. Repeating our analysis for more GCs would also
help to check for overall consistency, although the distance from
main-sequence fitting would suffer from common uncertainties caused
by the limited number of Hipparcos subdwarfs that can be used.

The stellar energy-loss limit remains the most restrictive constraint
on $\mu_\nu$. The most restrictive laboratory limit uses the
$\bar\nu_e$ flux from reactors and studies the electron recoil
spectrum upon $\bar\nu_e$ scattering, leading to the constraint
$\mu_{\bar\nu_e}<32\times10^{-12}\,\mu_{\rm B}$ (90\%~CL) on neutrino
magnetic or transition moments that are connected to $\bar\nu_e$
\cite{Beda:2009kx}. This quantity is different from our $\mu_\nu$,
which effectively sums over all direct and transition moments between
all flavors, and therefore is more general. It also applies to
transition moments between ordinary active and putative sterile
neutrinos, provided the latter are light enough to be emitted from
the degenerate helium core near the TRGB, i.e., the mass is safely
below the relevant plasma frequency of about 10--20~keV.

Globular clusters remain powerful---and in some cases
leading---particle physics laboratories. Their potential should be
fully exploited with contemporary observations and modern stellar
evolution theory.

%%%%%%%%%%%%%%%%%%%%%%%%%%%%%%%%%%%%%%%%%%%%%%%%%%%%%%%%%%%%%%%%%%%%%%
%Acknowledgements %%%%%%%%%%%%%%%%%%%%%%%%%%%%%%%%%%%%%%%%%%%%%%%%%%%%
%%%%%%%%%%%%%%%%%%%%%%%%%%%%%%%%%%%%%%%%%%%%%%%%%%%%%%%%%%%%%%%%%%%%%%

{\em Acknowledgments.}---Support for N.V.\ and M.C.\ is provided by
the Chilean Ministry for the Economy, Development, and Tourism's
Programa Iniciativa Cient\'{i}fica Milenio through grant P07-021-F,
awarded to The Milky Way Millennium Nucleus; by Proyecto Fondecyt
Regular \#1110326; by the BASAL Center for Astrophysics and
Associated Technologies (PFB-06); and by Proyecto Anillo ACT-86.
Support for N.V.\ is also provided by MECESUP Project No. PUC0609
(Chile). G.R.\ acknowledges partial support by the Deutsche
Forschungsgemeinschaft through grant EXC 153 and by the European
Union through the Initial Training Network ``Invisibles,'' grant
PITN-GA-2011-28944. J.R.\ acknowledges support by the Alexander von
Humboldt Foundation. A.A.R.V. acknowledge support from CNPq, CAPES, and INEspa\c{c}o agencies.

%%%%%%%%%%%%%%%%%%%%%%%%%%%%%%%%%%%%%%%%%%%%%%%%%%%%%%%%%%%%%%%%%%%%%%
%%%  Bibliography  %%%%%%%%%%%%%%%%%%%%%%%%%%%%%%%%%%%%%%%%%%%%%%%%%%%
%%%%%%%%%%%%%%%%%%%%%%%%%%%%%%%%%%%%%%%%%%%%%%%%%%%%%%%%%%%%%%%%%%%%%%

%%%%%%%%%%%%%%%%%%%%%%%%%%%%%%%%%%%%%%%%%%%%%%%%%%%%%%%%%%%%%%%%%%%%%%
\end{document}